\documentclass[journal]{IEEEtran}
\usepackage{graphicx}
\begin{document}
\title{Construction of Triple-GEM Detectors Using Commercially Manufactured Large GEM Foils}

\author{M.~Posik and
        B.~Surrow% <-this % stops a space
\thanks{Manuscript received November 30, 2016. This work was supported in part by an EIC grant, subtracted by Brookhaven Science Associates \#223228.}% <-this % stops a space
%\thanks{This work was supported in part by an EIC grant, subtracted by Brookhaven Science Associates #223228.}%
\thanks{M.~Posik is with Temple University, Philadelphia, PA 19122 USA (telephone: 215-204-8690, e-mail: posik@temple.edu).}%
\thanks{B.~Surrow is with Temple University, Philadelphia, PA 19122 USA (telephone: 215-204-7644, e-mail: surrow@temple.edu).}%
}

\maketitle
\pagestyle{empty}
\thispagestyle{empty}

\begin{abstract}
Many experiments are currently using or proposing to use large area GEM foils in their detectors, which is creating a need for commercially available GEM foils. Currently CERN is the only main distributor of large GEM foils, however with the growing interest in GEM technology keeping up with the increasing demand for GEMs will be difficult. Thus the commercialization of GEMs up to 50 $\times$ 50 cm$^2$ has been established by Tech-Etch Inc. of Plymouth, MA, USA using the single-mask technique.
The electrical performance and optical quality of the single-mask GEM foils have been found to be on par with those produced by CERN. The next critical step towards validating the Tech-Etch GEM foils is to test their performance under physics conditions. These measurements will allow us to quantify and compare the gain and efficiency of the detector to other triple-GEM detectors. This will be done by constructing several single-mask triple-GEM detectors, using foils manufactured by Tech-Etch, which follow the design used by the STAR Forward GEM Tracker (FGT). The stack is formed by gluing the foils to the frames and then gluing the frames together. The stack also includes a Tech-Etch produced high voltage foil and a 2D $r - \phi$ readout foil. While one of the triple-GEM detectors will be built identically to the STAR FGT, the others will investigate ways in which to further decrease the material budget and increase the efficiency of the detector by incorporating perforated Kapton spacer rings rather than G10 spacing grids to reduce the dead area of the detector.
The materials and tooling needed to assemble the triple-GEM detectors have been acquired. The GEM foils have been electrically tested, and a handful have been optically scanned. We found these results to be consistent with GEM foils produced by CERN. With the success of these initial tests, construction of the triple-GEM detectors is now under way.
\end{abstract}

%\begin{IEEEkeywords}
%IEEEtran, journal, \LaTeX, paper, template.
%\end{IEEEkeywords}

\section{Introduction}
\IEEEPARstart{M}{any} experiments in the nuclear and particle physics communities, such as STAR~\cite{Surrow:2010}, COMPASS~\cite{Altunbas:2002ds} and others, are currently employing the use of large-area GEM technology~\cite{Sauli:1997qp} in their detectors. With the maturity of GEM technology many future experiments, such as: the Super BigBite Spectrometer~\cite{SBS}, ALICE~\cite{Gasik:2014sga}, CMS~\cite{Abbaneo:2014}, Electron Ion Collider (EIC)~\cite{EIC}, and others, are proposing new detectors which incorporate large-area GEM technology. Such a demand for GEM foils will be difficult for CERN, who is currently the only primary distributor of large GEMs, to keep up with the demand. As a result, we had started a collaborative
effort with Tech-Etch Inc. of Plymouth, MA. to provide commercially
produced large-area GEM foils using single-mask etching techniques. The single mask process, developed by CERN~\cite{DuartePinto:2009yq}, allows for the production of large-area ($>$ 1 m long) GEM foils.

Before putting its GEM production on hold, Tech-Etch had successfully established a repeatable and stable procedure to manufacture GEM foils up to $50\times 50$ cm$^2$, which represents the first critical step to readily available large GEM foils.

In order to validate and ensure that the foils produced at Tech-Etch meet the needs of the general nuclear and particle physics communities, the quality of the foils need to be assessed. This is done in a two-step validation process. The first step involves measuring each foil's electrical and geometrical properties. These have been studied at Temple University using our optical analysis setup~\cite{Posik:2014,Becker:2006yc}. The single-mask Tech-Etch foils were found to be comparable to those produced at CERN and show excellent electrical and geometrical properties~\cite{Posik:2014}. The second validation step involves constructing a triple-GEM detector and measuring the overall gain of the detector, as well as its efficiency. These quantities can then be compared to previous or existing triple-GEM detectors to assess the overall quality of the commercially available Tech-Etch foils.

\section{Initial Validation}
Before any triple-GEM detectors can be built, the 40$\times$40 cm$^2$ single-mask GEM foils received from Tech-Etch (12 total) need to be electrically and optically tested. The electrical performance of all 12 Tech-Etch foils were assessed through leakage current measurements. The leakage current was found to be near or below 1 nA, which is expected given the resistivity of Apical~\cite{Posik:2014}. Optical analysis of three of the foils showed consistent and acceptable geometrical features, which include: inner hole diameter $\left(\approx\left<53\mu m\right>\right)$, outer hole $\left(\approx\left<78\mu m\right>\right)$diameter, and pitch $\left(\approx\left<138\mu m\right>\right)$. 

\section{Triple-GEM Detector Design}
\begin{figure}[!t]
\centering
\includegraphics[width=3.5in]{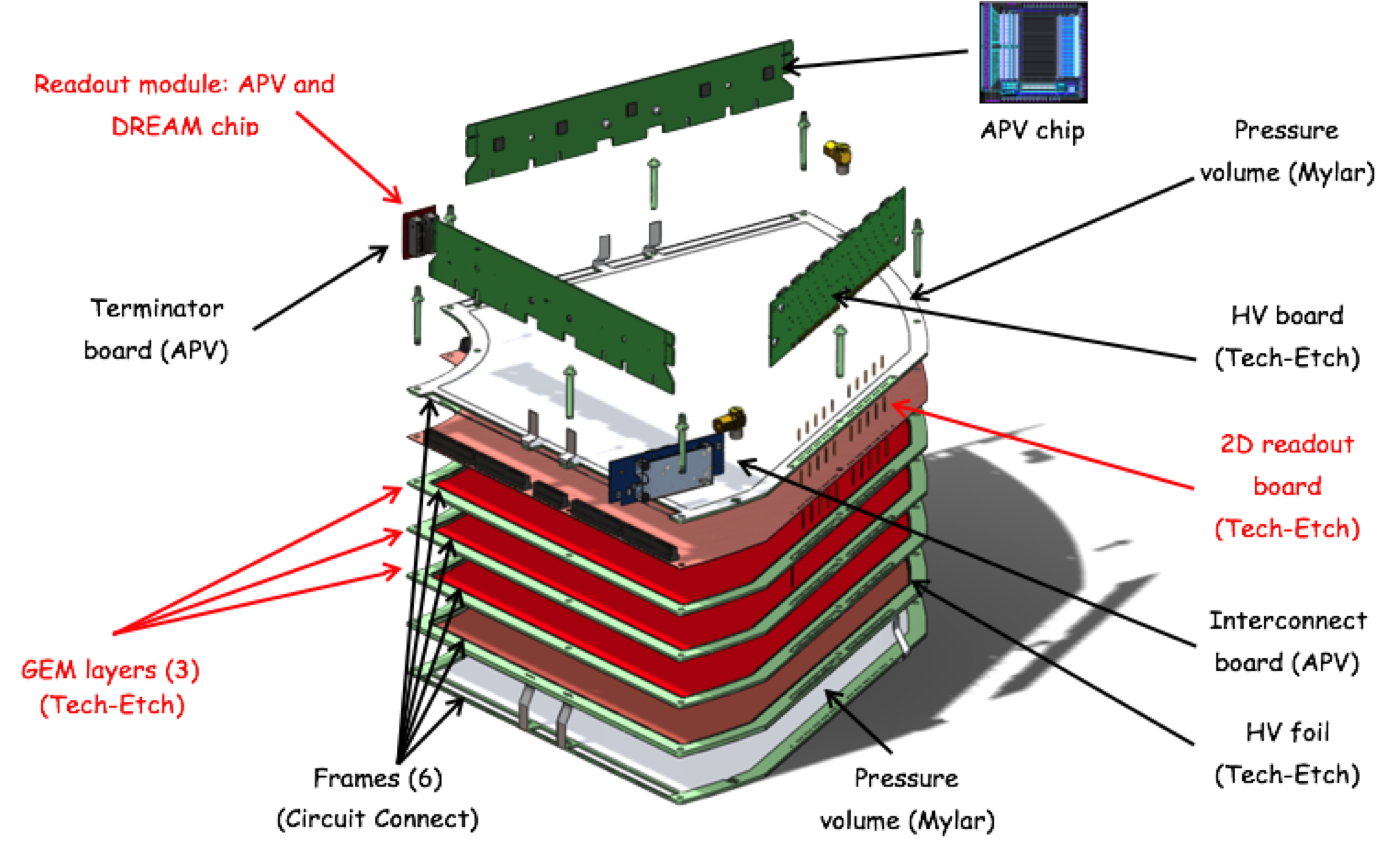}
\caption{Exploded view of triple-GEM detector.}
\label{fig_GEM-Exploded}
\end{figure}

Several triple-GEM detectors are being built using the Tech-Etch 40$\times$40 cm$^2$ single-mask foils to allow us to quantify the overall gain and efficiency of these commercial foils. To take advantage of existing tooling, designs, and testing equipment, the triple-GEM detectors will be built following the design of the STAR FGT~\cite{Surrow:2010}. The GEM stack consists of a 2 mm gap between the top pressure volume foil (PV) and high voltage foil (HV), 3 mm gap between the HV and first GEM foil (G1), a 2 mm gap between G1 and G2, a 2mm gap between G2 and G3, a 2mm gap between G3 and the readout foil (RO), and finally a 2 mm gap between the RO and bottom PV. The stack is assembled by gluing the foils to frames and then gluing the frames together. The triple-GEM volume will be filled with Ar-CO2 (90/10) gas mixture. The 40$\times$40 cm$^2$ single-mask GEMs, HV, and RO foils have all been produced by Tech-Etch. An exploded view of the triple-GEM detector is shown in Fig.~\ref{fig_GEM-Exploded} and details of the pieces and assembly will be given in the following sections. 

\subsection{GEM Foil}
\begin{figure}[!t]
\centering
\includegraphics[width=3.5in]{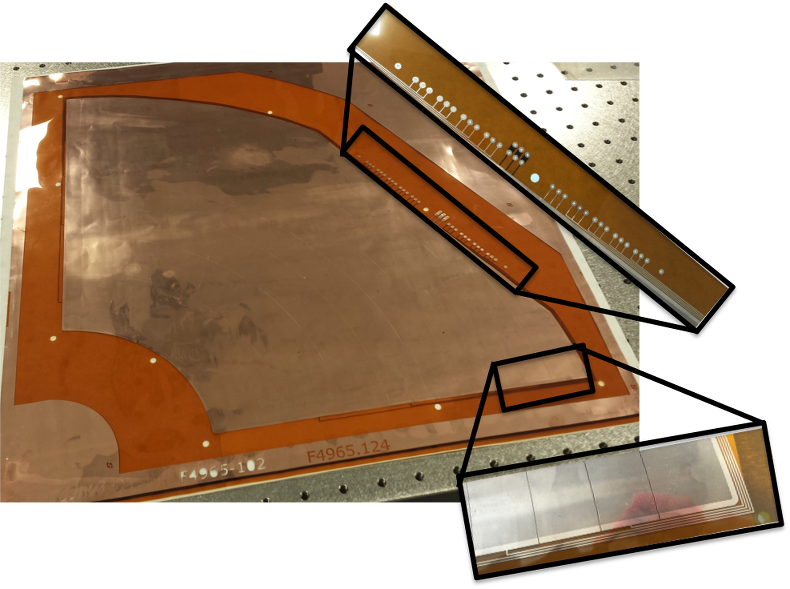}
\caption{Tech-Etch produced single-mask GEM foil.}
\label{fig_HV-foil}
\end{figure}

\begin{figure}[!t]
\centering
\includegraphics[width=3.0in]{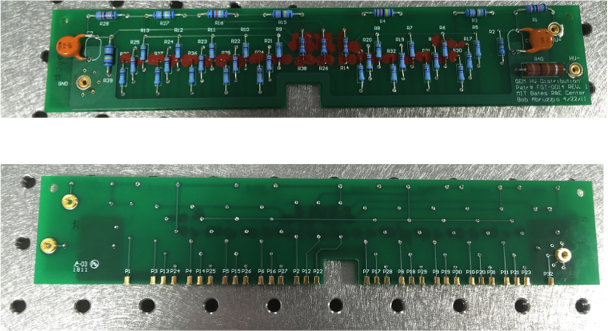}
\caption{HV distribution board front side (top image) and back side (bottom image).}
\label{fig_HV-Board}
\end{figure}

The GEM foil is a copper clad Apical foil. The front side of the foil is segmented into nine HV sections, each having an area of about 100 cm$^2$. Each of the high voltage segments are then routed along the side of the foil up to connectors at the top of the foil. Each of the HV segment routing lines are connected to nine sets of three connectors, which correspond to the nine HV segments for each of the three GEM foils in the triple-GEM stack. Figure~\ref{fig_HV-foil} shows one of the single-mask GEM foils and close up images of several routing lines and the HV connectors at the top of the foil. The three connectors at the top center of the foil (with the black band across them) connect to the backside of the foil, which is unsegmented and provides a ground for the foil. Each of the three unsegmented HV pins correspond to one of the three GEM layers. The HV will be distributed to the three GEM layers via a HV foil (also produced by Tech-Etch) and a HV board. An image of the HV board is shown in Fig.~\ref{fig_HV-Board}, where the various HV pins can be seen from the backside of the board. 

\subsection{Readout}
\begin{figure}[!t]
\centering
\includegraphics[width=3.0in]{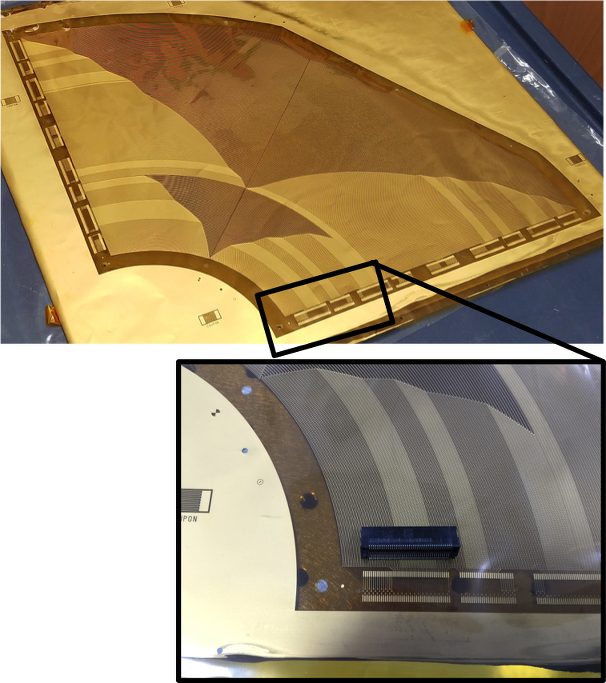}
\caption{Tech-Etch produced 2D readout foil and multi-pin connector.}
\label{fig_RO-Foil}
\end{figure}

The triple-GEM detectors will use the same Tech-Etch produced 2D readout foil design as the STAR FGT~\cite{Surrow:2010}. The top layer will read out the $\phi$ direction and the bottom layer the $r$ direction. The charges are readout from lines at constant angle, and connected pads at constant radius. The pitch design of the RO foils varies between about 300-900 $\mu m$ over the area of the foil. This variation in the pitch leads to the optical effect seen in Fig.~\ref{fig_RO-Foil} which shows an image of the RO foil. The 560 $r$ and 720 $\phi$ strips will be read out using two APV cards with each card containing ten APV-25 chips, giving a total of 1280 channels which will be connected via multi-pin connectors along the sides of the foil. The soldering of the multi-pin connectors is being carried out by Proxy Manufacturing Inc. The bottom image of Fig.~\ref{fig_RO-Foil} shows a close-up image of the multi-pin connector next to its connector pads on the RO foil. The APVs will then be read out using the same DAQ system that was designed for the STAR FGT detector, which is now located in the detector lab at Temple University. 

\subsection{Frames}
\begin{figure}[!t]
\centering
\includegraphics[width=3.0in]{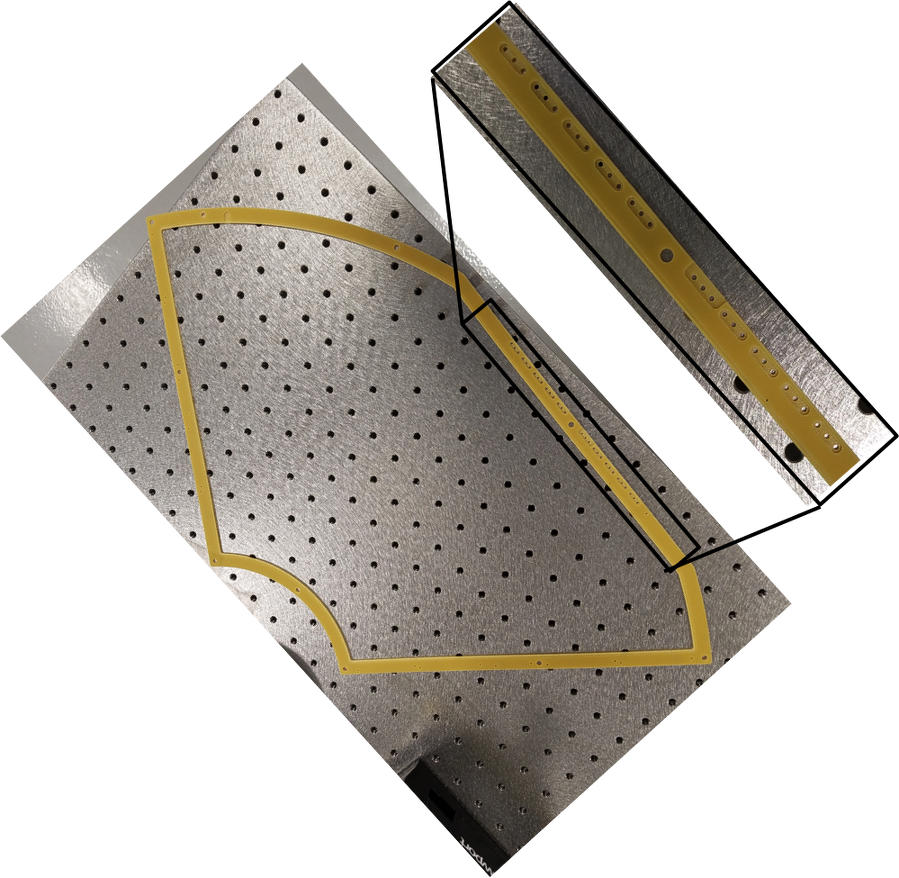}
\caption{2 mm spacer frame that is being used in the triple-GEM detector. The closeup image shows off the HV pin pass-through holes that allow the HV board to access each GEM layer.}
\label{fig_Frame}
\end{figure}

The frames that are being used for the triple-GEM detector were manufactured by Circuit Connect out of a FR4 rated VT-47 material and come in four varieties: top PV spacer, bottom PV spacer, 2 mm spacer, and 3 mm spacer. Each triple-GEM stack will have one top and bottom PV spacer, two 2 mm spacers, and one 3 mm spacer. Figure~\ref{fig_Frame} shows an image of one of the 2 mm spacer frames that is being used in the triple-GEM detector, along with a close-up image of the top of the frame which shows off the HV pin pass-through holes.    

\subsection{Foil Spacing}
\begin{figure}[!t]
\centering
\includegraphics[width=3.0in]{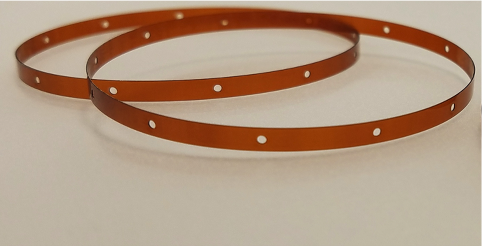}
\caption{Two Kapton rings that will be investigated as an alternative to using spacer grids.}
\label{fig_Ring}
\end{figure}

In a triple-GEM detector it is important to insure that GEM foil layers are not sagging or touching one another. To achieve this with the triple-GEM prototypes we will be implementing and testing the use of spacer rings (Fig.~\ref{fig_Ring}), rather than the more traditional FR4-G10 spacer grids. The spacer rings are perforated Kapton rings with an inner diameter of about 51 mm and wall thickness of 0.127 mm. The rings were cut from a tube of Kapton material into 2 and 3 mm high pieces. The cutting was done by Potomic Photonics using a laser cutting device. The perforated holes that were cut into the Kapton rings will allow gas flow through them. The Kapton rings have a couple of potential advantages over the spacer grid method. First the Kapton rings themselves have a lower material budget than the G10 spacer grids. Secondly, they could offer a cheaper alternative to space grids as the triple-GEM detectors become larger and approach a length of one meter. 

\begin{figure}[!t]
\centering
\includegraphics[width=3.0in]{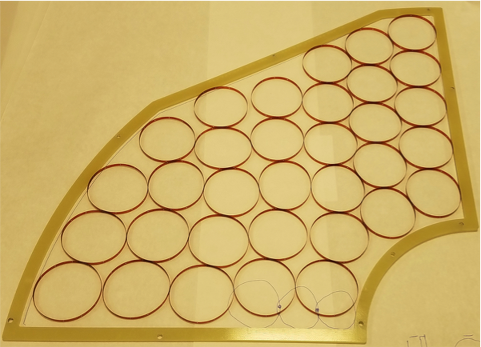}
\caption{Initial Kapton ring packing scheme to be tested in the first triple-GEM detector.}
\label{fig_Ring-Pack}
\end{figure}

There are many different packing arrangements that the Kapton rings can be implemented within the GEM volume. One initial packing arrangement that we will look at first is shown in Fig.~\ref{fig_Ring-Pack}. Using this packing scheme it appears that Kapton rings will not need to be glued and can simply be placed into the GEM volume as shown in Fig.~\ref{fig_Ring-Pack}.  

%\begin{figure*}[!t]
%\subfloat[Case I]\includegraphics[width=2.5in]{Figs/RO-Foil.png}%
%\label{fig_first_case}
%\hfil
%\subfloat[Case II]{\includegraphics[width=2.5in]{Figs/RO-Foil-Zoom.png}%
%\label{fig_second_case}}}
%\caption{Simulation results}
%\label{fig_sim}
%\end{figure*}

\section{Detector Assembly}
The assembly of the triple-GEM detectors will take place in Temple University's permanent class 1,000 clean room. The clean room facility has access to two types of gas, argon and nitrogen, in addition to a dry air line. We will make use of each of these three gasses through out the detector assembly process.

\begin{figure}[!t]
\centering
\includegraphics[width=3.0in]{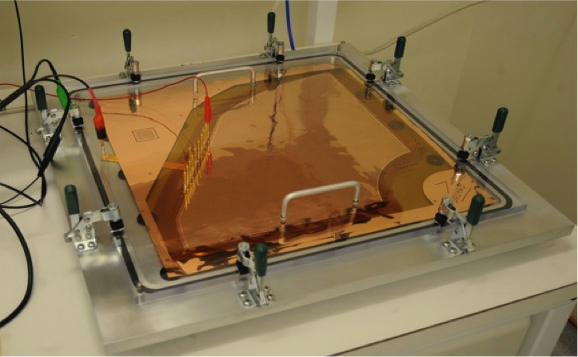}
\caption{Nitrogen enclosure which is used to carry out leakage current measurements of bare foils.}
\label{fig_N2-Box}
\end{figure}

 The detector assembly begins by first measuring the leakage current of the bare GEM foils. This is measured by placing each GEM foil into an nitrogen enclosure and applying up to 600 V to each HV segment in the foil and measuring the resulting current. An image of our nitrogen GEM enclosure can be seen in Fig.~\ref{fig_N2-Box}. All twelve 40$\times$40 cm$^2$ GEM foils that we received from Tech-Etch have now been measured and show leakage currents near or below 1 nA~\cite{Posik:2014}.

Once the bare GEM foils have been electrically tested, we then optically inspect them using our CCD scanner. This allows us to inspect each foil for hole and pitch uniformity. Currently we have optically analyzed three of the twelve 40$\times$40 cm$^2$ foils that we received from Tech-Etch~\cite{Posik:2014}. The remaining nine foils will be optically scanned once we have finished upgrading our scanning setup to accommodate foils up to one meter long. 

\begin{figure}[!t]
\centering
\includegraphics[width=3.0in]{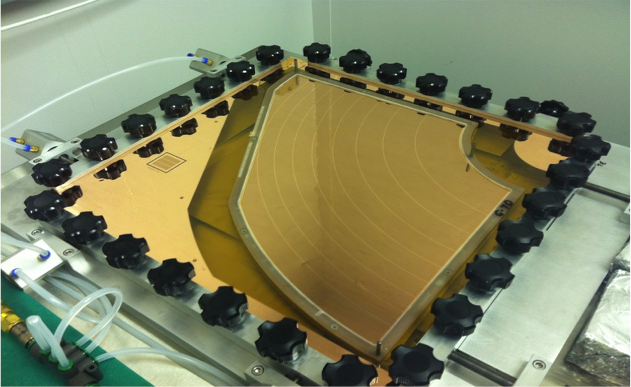}
\caption{A stretched GEM foil inside the GEM stretcher.}
\label{fig_Stretch}
\end{figure}

\begin{figure}[!t]
\centering
\includegraphics[width=3.0in]{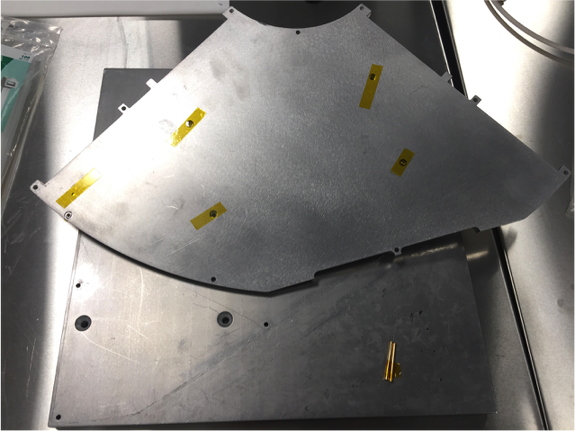}
\caption{Tooling used to apply pressure after the GEM foil is glued to the frame while it is drying. It has gas connections to allow flushing with gas to perform leakage current measurements.}
\label{fig_GEM-Dry}
\end{figure}

Next, the frames that will be used in the triple-GEM detector need to be cleaned. The cleaning will take place in a 55 L ultra sonic bath that is filled with de-ionized water. Once the frames are cleaned and have dried they will then be ready to start gluing the foils to their respective frames. This is done by placing the GEM foil into a GEM stretcher which uses four pressure cylinders (two per side) to apply tension to the GEM foil. The pressure to the cylinders is delivered via the dry air line in the clean room. The ripples and folds in the GEM foil are then worked out by manually adjusting the tension applied to the other two sides. Fig.~\ref{fig_Stretch} shows an image of a stretched GEM foil in the GEM stretcher. Once the foil is stretched and flat, glue (HT ARALDITE AY 103) will then be applied to the frame. The glue will be applied through a foot pedal dispensing system that ensures the glue is being steadily applied. The frame will then be glued to the foil. Once the foil is glued to the frame, the layer will be placed into the tooling shown in Fig.~\ref{fig_GEM-Dry} which applies pressure to assure good adhesion during drying. It also has gas connections to allow nitrogen gas to be flushed through the volume and have the foil electrically tested.

\begin{figure}[!t]
\centering
\includegraphics[width=3.0in]{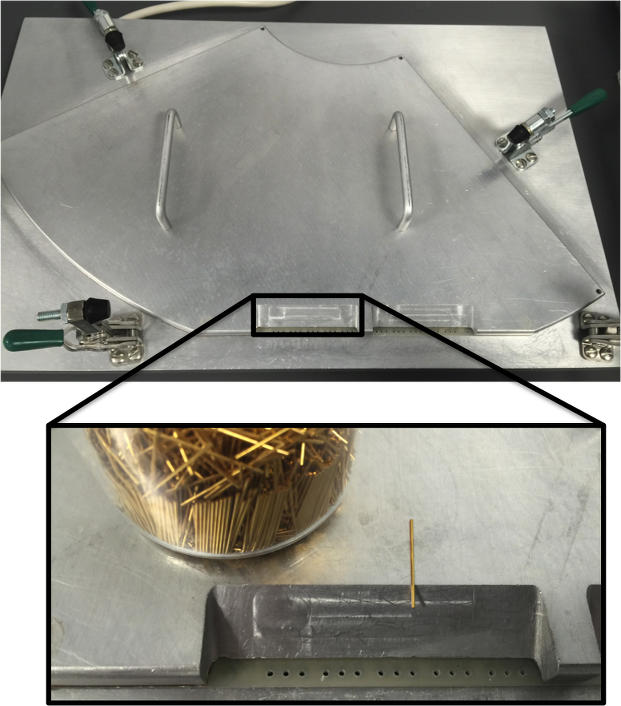}
\caption{Tooling used to solder HV pins to GEM HV connector pads.}
\label{fig_GEM-Solder}
\end{figure}

The GEM foil pad connections also need be soldered to HV pins that connect to the HV board. This will be done using the HV pin soldering tool shown in Fig.~\ref{fig_GEM-Solder}. This tooling allows the active area of the GEM to be protected and only exposes the GEM pad connections to which the HV pins are soldered. Once the GEMs are glued to the frames and HV pins are soldered to the GEM pad connectors, the triple-GEM stack will be formed by gluing the frames of each layer together. After each layer the detector will again be electrically tested. Once the PV layers are in place, the argon gas source will be used to check for leaks in the detector.

\section{Testing Setup}
\begin{figure}[!t]
\centering
\includegraphics[width=3.0in]{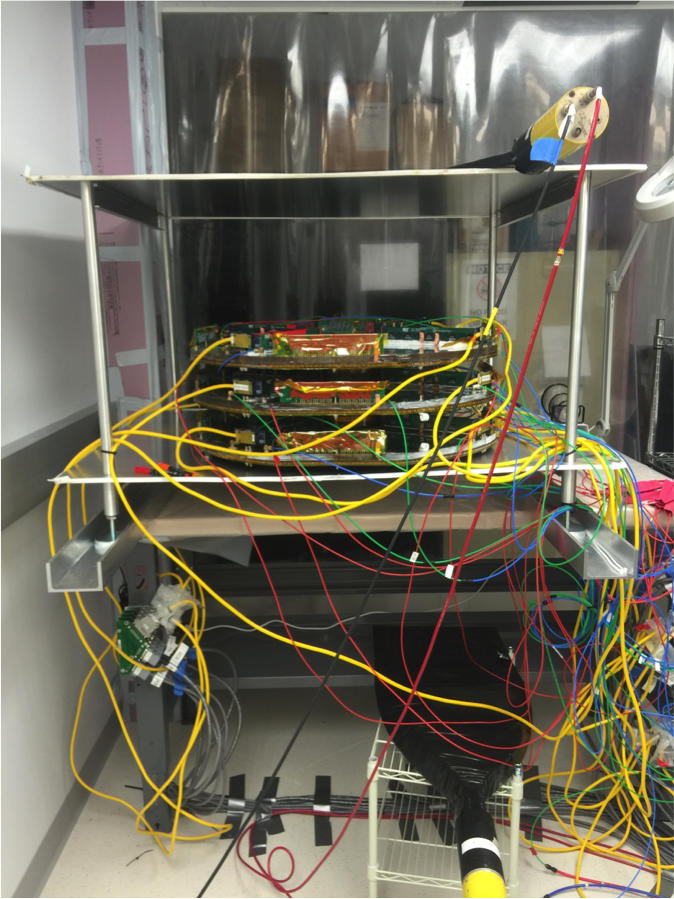}
\caption{Cosmic ray testing setup.}
\label{fig_Cosmics}
\end{figure}

Once the detector is assembled, it will be moved to Temple University's detector lab where the efficiency and gain will be measured. The efficiency will first be measured using our cosmic ray setup, seen in Fig.~\ref{fig_Cosmics} that shows part of the STAR FGT being tested with cosmics.

In addition to testing the efficiency with cosmic rays, we also plan on testing the energy resolution and gain via an $^{55}$Fe source. Complimentary to these measurements we also plan on using a Au target 50 keV mini X-ray tube to also measure the gain and efficiency. In order to look at the gain uniformity over the area of the triple-GEM detector, a X-Y scanner is being built that consists of two linear stages. This will allow us to localize the X-ray source to a particular area of the triple-GEM detector and study the gain uniformity as a function of the detector area.

\vfill

\section{Conclusion}

A new commercial source, within the US and in collaboration with Temple University, for large-area GEM foil fabrication based on single-mask etching techniques is currently being investigated. This new source would benefit from the knowledge acquired through Tech-Etch's successful commercialization of GEM foils (up to 50$\times$50 cm$^2$), in which Temple University played a vital role, and could provide the basis for application 
of large-area GEM foils by future experiments at CERN such as ALICE and CMS as 
well as upgrades at RHIC, JLAB, and a future EIC facility.

The first step towards validating the Tech-Etch GEM foils through electrical performance and optical analysis has yielded results consistent with what
is currently produced at CERN. The second validation step calls for characterizing the gain and efficiency of the GEM foils. As a result several triple-GEM detectors based on the Tech-Etch single-mask foils (40$\times$40 cm$^2$) are under construction. Additionally we are actively looking into ways in which to further reduce the material budget of a triple-GEM detector through the use of Kapton spacer rings to keep the GEM foil layers isolated from one another.

%\appendices
%\section{}
%Appendices, if needed, appear before the acknowledgment.

\section*{Acknowledgment}
We would like to thank David Crary, Kerry Kearney, and Matthew Campbell (Tech-Etch Inc.), as well as M. Hohlmann (FIT), R. Majka (Yale), and especially R. De Oliveira (CERN) for their useful discussions, guidance, and expertise which has lead to the successful commercialization of GEM technology. We would also like to thank D. Hasell, J. Kelsey, and J. Bessuille (MIT) for their help with designing, tooling, and assembly of the triple-GEM detectors.


%\bibliographystyle{IEEEtran}

\end{document}